\input epsf
\documentclass[12pt]{article}
\usepackage{fullpage}
\usepackage{graphicx}
\usepackage{pifont}

\def\n{{\rm n}}
\def\r{{\rm r}}
\def\s{{\rm s}}
\def\0{{\rm 0}}

\def\x{{\rm x}}

\def\i{{\rm i}}
\def\k{{\rm k}}

\def\gesim{\ \raise .6ex\hbox{$>$} \hskip -.11in\lower .5ex\hbox{$\sim$}\ }
\def\lesim{\ \raise .6ex\hbox{$<$} \hskip -.11in\lower .5ex\hbox{$\sim$}\ }

\begin{document}
\centerline{\bf Quantum Computation and Quantum Information:}
\centerline{\bf Are They Related to Quantum Paradoxology?}
\vskip .50truein
\centerline{Elias P. Gyftopoulos}
\centerline{Massachusetts Institute of Technology}
\centerline{Cambridge, Massachusetts  02139}
\vskip .50truein
\centerline{Michael R. von Spakovsky}
\centerline{Virginia Polytechnic Institute and State University}
\centerline{Blacksburg, Virginia 24061}
\vskip .50truein
\noindent
Submitted to PRB on Feb. 2, 2004.  Rejected by the editorial staffs of
both PRA and PRB on Feb. 3, 2004!
\vskip .50truein

We review both the Einstein, Podolsky, Rosen (EPR) paper about
the completeness of quantum theory, and Schr\"odinger's
responses to the EPR paper.  We find that both the EPR paper and
Schr\"odinger's responses, including the cat paradox, are 
not consistent
with the current understanding of quantum theory and thermodynamics.

\vskip .25truein

Because both the EPR paper and Schr\"odinger's responses play a
leading role in discussions of the fascinating and promising
fields of quantum computation and quantum information, we hope
our review will be helpful to researchers in these fields.

\vskip .50truein
\noindent
PACS number(s): 03.65.Bz, 03.65.-w, 05.70.Ln, 05.70.-a, Quantum
Computers.

\vfill\eject

\section{Introduction}

\hspace{.25truein} The great and very important interest in quantum computation and
quantum information [1] behooves us to review the current definitions,
postulates, and major theorems of quantum theory to see whether they
are correctly and consistently used by researchers in the development
of the fascinating and promising field of quantum computers.

We notice that practically all discussions of quantum computers and
quantum information involve the paradoxology of the famous Albert
Einstein, Boris Podolsky, and Nathan Rosen article [2], also known as
the EPR paper, and the Erwin Schr\"odinger publications that were
written in response to the EPR paper, and include his cat paradox
[3-6].

In this essay, we review the EPR and Schr\"odinger publications
just cited, and regret to report that they
misrepresent
the definitions, postulates, and principal theorems of quantum
theory.  The misrepresentations arise from: lack of clear and/or
complete definitions at an instant in time of the concepts system,
property, and state; use of a postulate that is proven to be
false; misinterpretation of uncertainty relations for observables 
that are described by non-commuting operators; oversight 
of the
fact that quantum -- nonstatistical -- probabilities are a linear
mapping of, and are solely determined by measurements of a complete
set of expectation values of linearly independent observables, and
conversely that a complete set of expectation values of linearly
independent observables are a linear mapping of the non-statistical 
probabilities.

We review also the concept of entanglement and find that its
interpretation in publications on quantum computation and quantum
information is based on the misconception 
that 
the expansion of a wave function in terms of a complete set
of orthonormal eigenfunctions is
a 
superposition of the eigenfunctions.

Our discussions are restricted to non-relativistic phenomena, and
are organized as follows.  We discuss the EPR paper in Section 2,
Schr\"odinger's publications and his cat paradox in Section 3, the
use
of the paradoxes in publications on quantum computation and quantum
information in Section 4, and our conclusions in Section 5.
We present the universal definitions of the concepts system, property, and
state, the quantum mechanical representations
of the concepts just cited, as well as a brief discussion of the
complete equation of motion of quantum theory in the Appendix.

Moreover, our discussions include extensive quotations from the
original publications so as to avoid misunderstandings.

\section{Can Quantum-Mechanical Description of Physical Reality be
Considered Complete?}

\subsection{Description by a wave function}

\hspace{.25truein}The question posed at 
the heading of this section is identical
with the title of the EPR paper.  In the abstract of the 
paper, the authors aver: 

\begin{quote}
\hspace{.25truein}``In a complete theory there is an element 
corresponding to each element of reality.  A sufficient condition
for the reality of a physical quantity is the possibility of 
predicting it with certainty, without disturbing the system. 
In quantum mechanics in the case of two physical quantities
described by non-commuting operators, the knowledge of one
precludes the knowledge of the other.  Then either (1) the description
of reality given by the wave function in quantum mechanics is not
complete
or (2) these two quantities cannot have simultaneous reality.
Consideration of the problem of making predictions concerning
a system on the basis of measurements made on another system
that had previously interacted with it leads to the result that
if (1) is false then (2) is also false.  One is thus led to
conclude that the description of reality as given by a wave
function is not complete.''
\end{quote}

The conclusion about a wave function is both false and correct.
As we discuss in the Appendix (A V), the description of the probabilities
of the physical reality represented by a wave function or a
projector 
$\mbox{\Pifont{psy}r}_\i = \mbox{\Pifont{psy}r}_\i^2$
is complete for phenomena that 
correspond to zero entropy physics and, therefore, the
sweeping conclusion just cited is not correct.

As we discuss also in the  Appendix (A III and A V), however, probabilities
associated with measurement results may require a representation
by a density operator 
$\mbox{\Pifont{psy}r}
> \mbox{\Pifont{psy}r}^2$ that involves no statistics
of the type introduced in either statistical classical 
mechanics or statistical
quantum mechanics.  In sharp contrast to the density operator defined
in statistical theories of physics, the density operator 
$\mbox{\Pifont{psy}r}$ just cited
involves only quantum probabilities or frequencies of measurement
results, and can be represented solely by a homogeneous ensemble, that
is, an ensemble of identical systems, identically prepared in
which {\it each member is characterized by the same density
operator $\mbox{\Pifont{psy}r} > \mbox{\Pifont{psy}r}^2$ as the ensemble.}
Therefore, the conclusion reached by EPR is correct but not for the
reason cited in their paper.

It is noteworthy that the concept of a homogeneous ensemble was
introduced by von~Neumann [7] only for wave functions or projectors.
Recognition that the concept  applies also to density operators that
involve no probabilities of statistical physics 
was realized by Hatsopoulos and
Gyftopoulos [8], and Jauch [9].  The density operators just cited
$(\mbox{\Pifont{psy}r} > \mbox{\Pifont{psy}r}^2)$ 
correspond to nonzero entropy physics, and therefore
the conclusion in the EPR paper about wave functions is correct.
Of course, the revolutionary discovery (in the sense described
by Kuhn [10]) about density operators that can be represented by
homogeneous ensembles was not known in the 1930's
(however, see also Section 3.2.2).

It is also noteworthy that the concept of entropy referred to
in the preceding comments differs from the numerous concepts
introduced in textbooks and scientific articles on quantum 
mechanics, thermodynamics, and statistical physics.  The 
new concept
is shown to be both a nonstatistical  intrinsic property of
any system 
(both macroscopic and microscopic, including one spin) 
in any state
(both thermodynamic equilibrium and not thermodynamic equilibrium),
and a measure of either the  quantum-mechanical
spatial shape of the constituents of the system [11,12],
or the orientation of spins within and on the Block sphere.

Next, we consider the EPR statement that 
\begin{quote}
``In quantum mechanics
in the case of two physical quantities described by non-commuting
operators, the knowledge of one precludes the knowledge of
the other''.
\end{quote}

We conclude that the statement 
is not correct.  We discuss the proof of
our conclusion in A XI and A XIV
where we show that the value
of an observable is an expectation value determined by an ensemble
of measurement results and not by the result of a single measurement.
So each observable has a value -- an expectation value -- regardless
of whether the operator that represents the observable commutes
or does not commute with operators of other observables.
Solid evidence for the remarks just cited is provided by the
Sterm-Gerlach and the Davisson-Germer experiments.

\subsection{Measurement of the values of two observables represented
by non-commuting operators}

\hspace{.25truein}The authors of the EPR 
paper base their conclusion that the
description of reality as given by a wave function is not
complete on the following considerations:

\begin{quote} 
\hspace{.25truein}``Whatever the meaning assigned to the
 term {\it complete},
the following requirement for a complete theory seems to be a
necessary one: {\it every element of the physical reality
must have a counterpart in the physical theory.}  We shall call
this the condition of completeness. ...
The elements of the physical reality cannot be determined by 
{\it a priori} philosophical considerations, but must be found
by an appeal to results of experiments and measurements.  A
comprehensive definition of reality is, however, unnecessary
for our purpose.  We shall be satisfied with the following
criterion, which we regard as reasonable. 
{\it If, without in any way disturbing a system, we can
predict with certainty (i.e., with probability equal to
unity) the value of a physical quantity, then there exists
an element of physical reality corresponding to this
physical quantity.} ...
Regarded not as a necessary, but merely as a sufficient,
condition of reality, this criterion is in
agreement with classical as well as quantum-mechanical
ideas of reality''.
\end{quote}

The statements just cited are excellent and consistent with
the thoughts of many preeminent physicists, including the
description of ``The Nature of Physical Reality'' by Margenau~[13].

Next, the authors of the EPR paper assert:

\begin{quote}
\hspace{.25truein}``To illustrate the ideas involved let us consider the
quantum-mechanical description of the behavior of a particle
having a single degree of freedom.  The fundamental concept of
the theory is the concept of {\it state}, which is supposed to
be completely characterized by the wave function $\psi$, which
is a function of the variables chosen to describe the particle's
behavior.  Corresponding to each physically observable quantity
$A$ there is an operator, which may be designated by the same
letter.''
\end{quote}

We find this assertion misconceived because the wave function
does not determine the state.  As we discuss in the Appendix,
for all paradigms of physics
at an instant in time the definition
of state requires the definitions of both a system (A I) -- types
and amounts of constituents, inter-constituent or internal forces,
and external forces or parameters -- as an entity separable
from and uncorrelated with its environment, and the values of
a complete set of linearly independent properties
(A VI, A VII, and A XIV).  Overlooking
these definitions results in conclusions that misrepresent the
powerful and successful theory of quantum physics.

Another way of expressing the requirements just cited is to say that
given a system at an instant in time, its state is defined either by
a complete set of expectation values of linearly independent
properties -- observables -- or by both a complete set of linearly
independent operators representing observables, and either a wave
function or projector
$(\mbox{\Pifont{psy}r}_\i = \mbox{\Pifont{psy}r}_\i^2)$, 
or a density operator
$(\mbox{\Pifont{psy}r}> \mbox{\Pifont{psy}r}^2)$,
 where linear independence means that if an observable
represented by operator A is included in the list of
observables, then A$^\n$ for $\n > 1$ is
excluded.

The clarifications just cited are important for at least two
reasons: (i) Systems with different constituents, such as an
electron, or a helium atom, may have identical probability
characteristics that correspond to the same 
projector -- wave function -- or
density operator even though each of the two systems has a
different set of operators of quantum observables; and (ii) The
universal
definition of state for all paradigms of physics involves both
the concept of system, and the concept of property, both at an
instant in time (see A I, A VI, and A XII).

Next, the authors of the EPR paper describe the following example:

\begin{quote}
\hspace{.25truein}``If $\psi$ is an eigenfunction of the operator $A$, that is, if
\begin{equation}
\psi' \equiv A\psi = a\psi,
\end{equation}
where $a$ is a number, then the physical quantity
$A$ has with certainty the value $a$ whenever the particle
is in the state given by $\psi$.  In accordance with our
criterion of reality, for a particle in the state given by
$\psi$ for which Eq. (1) holds, there is an element of
physical reality corresponding to the physical quantity $A$.
Let, for example,
\begin{equation}
\psi = e^{\left( 2 \pi i/h \right) p_0x},
\end{equation}
where $h$ is Planck's constant, $p_0$ is some constant
number, and $x$ the independent variable.  Since the
operator corresponding to the momentum of the particle is
\begin{equation}
p = (h/2 \pi i) \partial/\partial x,
\end{equation}
we obtain
\begin{equation}
\psi' = p\psi = ( h/2\pi i) \partial \psi/\partial x = 
p_0 \psi.
\end{equation}
Thus, in the state given by Eq. (2), the momentum has
certainly the value $p_0$.  It thus has meaning to say that
the particle in the state given by Eq. (2) is real.''

\hspace{.25truein}``On the other hand if Eq. (1) does not hold, we can no longer
speak of the physical quantity $A$ having a particular value.
This is the case, for example, with the coordinate of the
particle.  The operator corresponding to it, say $q$, is the
operator of multiplication by the independent variable.
Thus,
\begin{equation}
q\psi = x\psi \not= a\psi.
\end{equation}
In accordance with quantum mechanics we can only say
that the relative probability that a measurement of
the coordinate will give a result lying between
$a$ and $b$ is
\begin{equation}
P(a,b) = \int_a^b \bar\psi \psi dx = \int_a^b dx = b -a.
\end{equation}
Since this probability is independent of $a$, but depends
only upon the difference $b-a$, we see that all values of
the coordinate are equally probable.''

\hspace{.25truein}``A definite value of the coordinate, for a particle in the
state given in Eq. (2), is thus not predictable, but may be
obtained only by a direct measurement.  Such a measurement
however disturbs the particle and thus alters its state.
After the coordinate is determined, the particle will no
longer be in the state given by Eq. (2).  The usual conclusion
from this in quantum mechanics is that {\it when the momentum
of a particle is known, its coordinate has no physical reality.}''
\end{quote}

The conclusions derived from the preceding  
example misrepresent the science of quantum mechanics for the
following
reasons: (i) If a momentum measurement yields the value $p_0$, that
result does not necessarily mean the system immediately after the
measurement is in a state for which the probabilities are described
by the wave function shown in Eq. (2).  In principle, an infinite
number of measurements, on identical systems, identically
prepared, all yielding the same result $p_0$ is necessary to ascertain
that the wave function is given by Eq. (2)
(see A X to A XIV); (ii) If the result of
a measurement is $p_0$, this does not mean that the value of the
momentum of the system -- ensemble member -- after the measurement is
$p_0$.  The conclusion that the momentum immediately after the
measurement
is $p_0$ is based on the so-called von Neumann's projection or
collapse
of the wave function postulate which 
is proven not to be valid [14-17, A XVI];
and (iii) If the expectation value of
momentum measurement results is indeed the $p_0$ of Eq. (2), then this does
not mean that the coordinate of the particle has no physical
meaning.  It simply means that the standard deviation of momentum
measurement results $\Delta p = 0$ and the standard deviation
of position measurement results  $\Delta x = \infty$ and such that
$\Delta x \Delta p = \infty \times 0 \geq \hbar/2$.  On the other
hand, if the particle is confined within an one-dimensional infinitely
deep potential well of width $L$, then the standard deviation
of position measurement results $\Delta x < L$, and the wave
function is not and cannot be given by Eq. (2) because, if it were, then
$\Delta p = 0$ and the uncertainty relation is violated, that is,
$\Delta x \Delta p = L \times 0 = 0$ and not $\geq \hbar/2$.

Next, the authors of the EPR paper aver: 

\begin{quote}
\hspace{.25truein}``In quantum mechanics
it is usually assumed that the wave function {\it does} contain a
complete description of the physical reality of the system in the
state to which it corresponds.  At first sight this assumption is
entirely reasonable, for the information obtainable from a wave
function
seems to correspond exactly to what can be measured without altering
the state of the system.  We shall show, however, that this
assumption,
together with the criterion of reality given above, leads to
a contradiction.''

\hspace{.25truein}``For this purpose let us suppose that we have two systems,
I and II, which we permit to interact from the time $t=0$ to
$t=T$, after which time we suppose that there is no longer any
interaction between the two parts.  We suppose further that the states
of the two systems before $t=0$ were known.
We can then calculate with the help of Schr\"odinger's equation
the state of the combined system I + II at any subsequent time;
in particular, for any $t>T$.  Let us designate the corresponding
wave function by $\Psi$.  We cannot, however, calculate the state in
which either one of the two systems is left after the interaction.
This, according to quantum mechanics, can be done only with the
help of further measurements, by a process known as the
{\it reduction of the wave packet.}  Let us consider the essentials
of this process.''

\hspace{.25truein}``Let $a_1, a_2, a_3, ...$ be the eigenvalues of some physical
quantity $A$ pertaining to system I and $u_1(x_1)$, 
$u_2(x_1)$, $u_3(x_1), ...$ the corresponding eigenfunctions,
where $x_1$ stands for the variables used to describe the first
system.  Then $\Psi$, considered as a function of $x_1$, can be 
expressed as:
\begin{equation}
\Psi \left( x_1,x_2 \right) = \sum_{\n=1}^\infty \psi_\n
\left( x_2 \right) u_\n \left( x_1 \right) ,
\end{equation}
where $x_2$ stands for the variables used to describe
the second system.  Here $\psi_n(x_2)$ are to be regarded
merely as the coefficients of the expansion of $\Psi$ into a
series of orthogonal functions $u_{\bf n}(x_1)$.
Suppose now that the quantity $A$ is measured and it is found that
it has the value $a_\k$.  It is then concluded that after the
measurement the first system is left in the state given by
the wave function $u_\k(x_1)$, and that the second system is
left in the state given by the wave function $\psi_\k(x_2)$.  This 
is the process of reduction of the wave packet; the wave packet
given by the infinite series (7) is reduced to a single term
$\psi_\k (x_2) u_\k (x_1)$.''

\hspace{.25truein}``The set of functions $u_\n(x_1)$ is determined by the choice
of physical quantity $A$.  If, instead of this, we had chosen another
quantity, say $B$, having the eigenvalues $b_1, b_2, b_3, ...$ and
eigenfunctions $v_1(x_1), v_2(x_1), v_3(x_1), ...$ we should
have obtained, instead of Eq. (7), the expansion
\begin{equation}
\Psi \left( x_1,x_2 \right) = \sum_{\s=1}^\infty 
\varphi_\s \left( x_2 \right) v_\s \left( x_1 \right) ,
\end{equation}
where $\varphi_\s$'s are the new coefficients.  If now the
quantity $B$ is measured and is found to have the value
$b_\r$, we conclude that after the measurement the first system
is left in the state given by $v_\r(x_1)$ and the second
system is left in the state given by $\varphi_\r(x_2)$.''

\hspace{.25truein}``We see therefore that, as a consequence of two different
measurements performed upon the first system, the second system
may be left in states with two different wave functions.  On the
other hand, since at the time of measurement the two systems no
longer interact, no real change can take place in the second
system in consequence of anything that may be done to the first
system.  This is, of course, merely a statement of what is meant
by the absence of an interaction between the two systems.  
Thus, {\it it is possible to assign two different wave functions}
(in our example $\psi_\k$ and $\varphi_\r)$ 
{\it to the same reality} (the second system after the interaction
with the first).''

\hspace{.25truein}``Now, it may happen that the two wave functions, $\psi_\k$ and
$\varphi_\r$, are eigenfunctions of two non-commuting operators
corresponding to some physical quantities $P$ and $Q$
respectively.  That this may actually be the case can best be
shown by an example.  Let us suppose that the two systems are two
particles, and that
\begin{equation}
\Psi \left( x_1, x_2 \right) = \int_{-\infty}^\infty e^
{\left( 2\pi i/h \right) \left( x_1-x_2+x_0 \right)p}
dp ,
\end{equation}
where $x_0$ is some constant.  Let $A$ be the momentum of the
first particle; then, as we have seen in Eq. (4), its
eigenfunctions will be
\begin{equation}
u_p \left( x_1 \right) = e^{\left( 2\pi i/h \right) px_1}
\end{equation}
corresponding to the eigenvalue $p$.  Since we have here the
case of a continuous spectrum, Eq. (7) will now be written
\begin{equation}
\Psi \left( x_1,x_2 \right) = \int_{-\infty}^\infty 
\psi_p \left( x_2 \right) u_p \left( x_1 \right) dp ,
\end{equation}
where
\begin{equation}
\psi_p \left( x_2 \right) = e^{- \left( 2\pi i/h \right)
\left( x_2-x_0 \right)p} .
\end{equation}
This $\psi_p$, however, is the eigenfunction of the operator
\begin{equation}
P = \left( h/2\pi i \right) \partial/\partial x_2 ,
\end{equation}
corresponding to the eigenvalue $-p$ of the momentum of the
second particle.  On the other hand, if $B$ is the coordinate
of the first particle, it has for eigenfunctions
\begin{equation}
v_\x \left( x_1 \right) = \delta \left( x_1 - x \right) .
\end{equation}
corresponding to the eigenvalue $x$, where 
$\delta (x_1-x)$ is the well-known Dirac delta-function.
Eq. (8) in this case becomes
\begin{equation}
\Psi \left( x_1,x_2 \right) = \int_{-\infty}^\infty 
\varphi_\x \left( x_2 \right) v_\x \left( x_1 \right) 
dx ,
\end{equation}
where
\begin{equation}
\varphi_\x \left( x_2 \right) = \int_{-\infty}^\infty 
e^{\left( 2\pi i/h \right) \left( x-x_2+x_0 \right) p}
dp = h \delta \left( x-x_2+x_0 \right) .
\end{equation}
This $\varphi_\x$, however, is the eigenfunction of the operator
\begin{equation}
Q = x_2
\end{equation}
corresponding to the eigenvalue $x+x_0$ of the coordinate
of the second particle.  Since
\begin{equation}
PQ - QP = h/2\pi i ,
\end{equation}
we have shown that it is in general possible for $\psi_\k$
and $\psi_\r$ to be eigenfunctions of two noncommuting operators,
corresponding to physical quantities.''

\hspace{.25truein}``Returning now to the general case contemplated in Eqs. (7) 
and (8), we assume that $\psi_\k$ and $\varphi_\r$ are
indeed eigenfunctions of some noncommuting operators $P$
and $Q$, corresponding to the eigenvalues 
$p_\k$ and $q_\r$, respectively.  Thus, by measuring either
$A$ or $B$ we are in a position to redirect with certainty,
and without in any way disturbing the second system, either the
value of the quantity $P$ (that is $p_\k$) or the value of the
quantity $Q$ (that is $q_\r$).  In accordance with
our criterion of reality, in the first case we must consider
the quantity $P$ as being an element of reality, in the second case
the quantity $Q$ is an element of reality.  But, as we have seen,
both wave functions $\psi_\k$ and $\varphi_\r$ belong to the
same reality.''
\end{quote}

In general, the calculations for $t>T$ are not
correct for several reasons: (i) For example, assume that
for $t \leq 0$ system I is a proton in a box, and system II
an electron in a box, and that during the interaction from
$t = 0$ to $t = T$ the two particles combine and form a hydrogen
atom.  The Hamiltonian operator after the interaction includes
the potential energy between the proton and the electron, an operator
that is absent from the Hamiltonian operator of the proton, and 
the Hamiltonian operator of the electron.  As a result, there are
no two systems that can be identified after the interaction; 
(ii) As we discuss earlier, the reduction of the wave packet 
is proven to be invalid.  Accordingly, the conclusion derived from
Eqs. (7-18) is not valid; and (iii) A third and very important
reason is that after the interaction is over, the two parts may be
separable but correlated, or what Schr\"odinger characterizes as
the systems are ``entangled'' [4].  If this is the case, defining
two systems 
after the interaction has ceased
amounts to neglect of correlations.  But neglect of correlations
results in an increase of the non-statistical entropy. 
A simple proof of the assertion just cited is provided. 
by the quantum interpretation of the Boltzmann equation and its 
collision integral.  An important consequence of this omission is
that any wave function $\Psi$ and its unitary evolution in time --
Schr\"odinger's equation of motion -- are not consistent with the
assumption that there are two systems after the separation at time
$T$ because this assumption is tantamount to an entropy greater than
zero, whereas any $\Psi$ corresponds to zero entropy.
Because every system in any state has entropy as a fundamental
and intrinsic property of the constituents (see A V), entropy
should not be created and/or destroyed by inappropriate
mathematical representations.

\subsection{Summary of comments on the EPR paper}

\hspace{.25truein}Many unwarranted conclusions 
in the EPR paper
mar the discussion of two systems
before and after an interaction between them.  They arise from not
using
clear definitions of the concepts of system, property, and
state, in that order; from the use of the
invalid  postulate of collapse
of a wave function as a result of a measurement; from  the omission
of radical effects that interactions may have on two initially
identifiable systems; and finally and most importantly 
from the oversight
of the fact that, in principle, in any problem involving 
either statistical or quantum probabilities, an infinite
number of measurements is required in order to establish
either the probability distribution, or the probability density
function, or a combination of these two, or more importantly
for quantum theory, the value -- the expectation value --
of any observable.

\section{The Present Situation in Quantum Mechanics}

\subsection{Schr\"odinger's response to the EPR paper}

\hspace{.25truein}The title of this section is a translation of the title of
three articles that Schr\"odinger authored in German [3], and
that have been translated into English [4].  These articles
include a one paragraph description of the ``cat paradox''.  In
addition to these articles, two presentations were made on
Schr\"odinger's behalf at the Cambridge Philosophical Society
by Born [5], and Dirac [6].  The title of both presentations is
``Discussion of Probability Relations between Separated Systems''.
In both Refs. [4] and [5], Schr\"odinger acknowledges that the EPR
paper motivated his offerings.

Schr\"odinger begins his discussion of quantum mechanics with the
following statements:

\begin{quote}
\hspace{.25truein}``{\it Statistics of Model Variables in Quantum Mechanics.}
At the pivot point of contemporary quantum mechanics (Q.M.) stands
a doctrine, that perhaps may yet undergo many shifts of meaning
but that will not, I am convinced, cease to be the pivot point.
It is this, that models with determining parts that uniquely 
determine each other, as do the classical ones, cannot do justice 
in nature.''
``One might think that for anyone believing this, the classical
models have played out their roles.  But this is not the case.''

\hspace{.25truein}``$A.$ The classical 
concept of state becomes lost, in that at most
a well-chosen {\it half} of a complete set of variables can be
assigned definite numerical values; ... The other half then
remains completely indeterminate, while supernumerary parts can show
highly varying degrees of indeterminacy.  In general, of a complete
set ... 
{\it all} will be known only uncertainly.  One can best keep track of
the degree of uncertainty by following classical mechanics and
choosing
variables arranged {\it in pairs} of so-called canonically-conjugate 
ones. The simplest example is a space coordinate x of a point mass
and the component p$_\x$ along the same direction, its linear 
momentum (i.e, mass times velocity).  Two such constrain each other
in the precision with which they may be simultaneously known, in
that the product of their tolerance -- or variation-widths
(customarily designated by putting a $\Delta$ ahead of the quantity)
cannot fall {\it below} the magnitude of a certain universal constant,
thus
\begin{equation}
\Delta \x \cdot \Delta {\rm p}_\x \underline{\geq} 
\ \ ({\rm \ Planck's \ constant)} / 2\pi = {\rm \hbar} . \nonumber
\end{equation}
\end{quote}

As we discuss in the Appendix, the conclusions just cited are
not correct because they overlook both the universal
definition of state (A XIV), and the meaning of uncertainty relations
(A XV and [1, 9]).

Next, Schr\"odinger states:

\begin{quote}
\hspace{.25truein}``$B.$ If even at any given moment not all variables are determined
by some of them, then of course neither are they all determined
for a later moment by data obtainable earlier.  This may be called 
a break with causality, but in view of $A.$ it is nothing
essentially new.  If a classical state does not exist at any moment,
it can hardly change causally.  What do change are the 
{\it statistics} or
{\it probabilities}, these moreover causally.  Individual variables
meanwhile may become more, or less, uncertain.  Overall it may
be said that the total precision of the description does not
change with time, because the principle of limitations described
under $A.$ remains the same at every moment.''
\end{quote}

As we discuss in Section 2.2, these remarks are not consistent
with either the equation of motion of quantum theory, or the 
definition of state.  What defines the state at an instant in
time is a set of expectation values.  If we restrict our
considerations
to probabilities that are described by a projector, then for any
expectation value $<$F$>$ it is readily shown that~[18]
\begin{displaymath}
 { d <{\rm F}> \over dt } = { \i \over {\rm \hbar}} < {\rm HF - FH} > 
+ { \partial <{\rm F}> \over \partial t}  
\end{displaymath}
where H is the Hamiltonian operator of the system.

Next, Schr\"odinger raises the question ``Can one base the
theory on ideal ensembles?'', and responds as follows.

\begin{quote}
\hspace{.25truein}``The classical model plays a Protean role in Q.M.  Each
of its determining parts can under certain circumstances become
an object of interest and achieve a certain reality.  But never
all of them together -- now it is these, now those, and indeed
always at most {\it half} of the complete set of variables allowed
by a full picture of the momentary state.  Meantime, how about
the others?  Have they then no reality, perhaps (pardon the
expression) a blurred reality; or are all of them always real and
is it merely, according to Theorem $A.$ that simultaneous 
{\it knowledge}
of them is ruled out?''
\end{quote}

As we discuss earlier, this dilemma does not exist if the proper
interpretation of quantum theory is followed.

Next, Schr\"odinger elaborates on the issue of blurred variables,
and introduces his cat paradox.

\begin{quote}
\hspace{.25truein}``One can even set up quite ridiculous cases.  A cat is penned up in
a steel chamber, along with the following diabolical device 
(which must be secured against direct interference by the cat): in
a Geiger counter there is a tiny bit of radioactive substance, so
small, that perhaps in the course of one hour one of the atoms
decays, but also, with equal probability, perhaps none; if it
happens, the counter tube discharges and through a relay
releases a hammer which shatters a small flask of hydrocyanic acid.
If one has left this entire system to itself for an hour, one would
say that the cat still lives if meanwhile no atom has decayed.
The first atomic decay would have poisoned it.  The $\psi$-function
of the entire system would express this by having in it the living
and the dead cat (pardon the expression) mixed or smeared out in
equal parts.''

\hspace{.25truein}``It is typical of these cases that an indeterminacy originally 
restricted to the atomic domain becomes transformed into
macroscopic indeterminacy, which can then be resolved by direct
observation.  That prevents us from so naively accepting as valid a
``blurred model'' for representing reality.  In itself it would not
embody anything unclear or contradictory.  There is a difference
between a shaky or out-of-focus photograph and a snapshot of
clouds and fog banks.''
\end{quote}

Upon using the definitions, postulates, and major theorems of
quantum theory
as outlined in the Appendix,
we conclude that the ``cat paradox''
represents no physical reality, that is, there exists no cat
paradox.  To facilitate our discussion, we use a cartoon (Figure 1)
that appears in a publication on quantum entanglement [19],
and that correctly claims to represent the cat paradox specified by
Schr\"odinger.

\begin{figure}[here]
\smallskip
\centerline{\includegraphics[width=4truein]{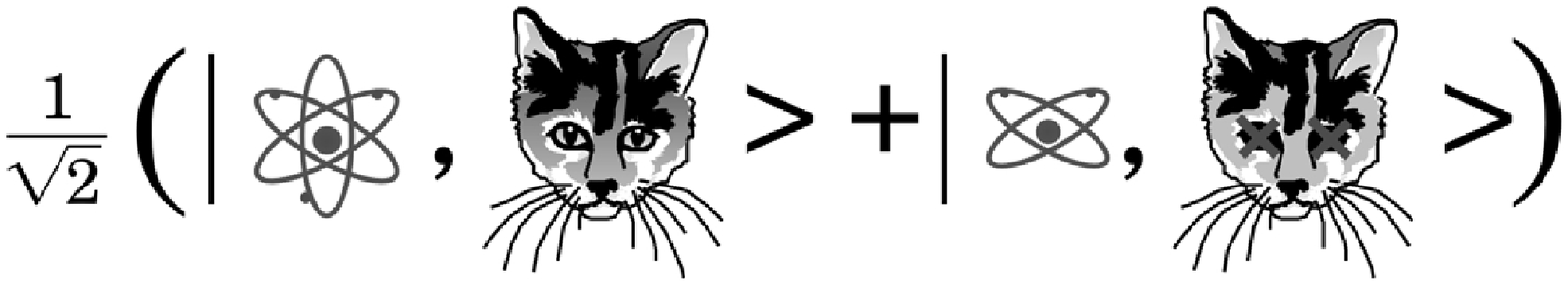}}
\caption{Cartoon representing Schr\"odinger's cat paradox.
(Reprinted from Physics Today by permission of the authors [19],
and Physics Today.)}
\end{figure}

The cartoon shows a ket that presumably can be 
represented by a superposition of and not by an expansion
in terms of 
two kets,
one consisting of a radioactivity source and a live cat at an 
instant in time $t$, and the other a radioactivity source that has decayed
at time $t+\tau$ and a dead cat that has been poisoned by the release
of hydrocyanic acid induced by the radiation emitted at $t+\tau$.

The cartoon represents accurately Schr\"odinger's cat paradox but
corresponds to no physical reality
because a ket is valid at a specific
instant in time, and therefore cannot be represented by 
an expansion in terms of 
let alone superposition of  two kets, one of which
applies at time $t$, and the other at time $t+\tau$.  Moreover,
and for sure more importantly, a radioactivity source prior to
decay is a system, that is, an entity both separable from and
uncorrelated with its environment which includes a live cat and its
life support interactions, such as breathing, drinking, eating, and
(excuse the expression) other necessities of living beings.  Solid and
incontrovertible evidence for the physical realities just cited is
provided by a very large number of radioactivity sources 
devoid of evil contraptions in hospitals, science and engineering
laboratories, and nuclear energy installations, and a myriad of 
creatures, including human beings and cats, that live happily around
these sources.

For clarity and avoidance of misinterpretations, at time $t$ the
Hilbert space of the radioactivity source and the cat must be the
direct
product $\mathcal{H}_{\r1}\otimes\mathcal{H}_{{\rm lc}}$, and the
catalog of probabilities by the direct product 
$\mbox{\Pifont{psy}r}_{\r1}\otimes 
\mbox{\Pifont{psy}r}{_{\rm lc}}$,
where $\mathcal{H}_\i$ and 
$\mbox{\Pifont{psy}r}_\i$ for i=r1, lc  are the Hilbert spaces
and 
probability catalogs 
of the radioactivity source and the live cat,
respectively.  Most likely, even though not necessarily, 
the 
probability catalogs are 
density operators for both the radioactivity source and the live 
cat and not projectors or, equivalently, the entropy of each of these
two systems is not zero.  Moreover, each of the 
probability catalogs
$\mbox{\Pifont{psy}r}_{\r1}$ and $\mbox{\Pifont{psy}r}_{{\rm lc}}$ is 
represented by a homogeneous or irreducible
ensemble of identical systems, identically prepared
(see A III and A V).
It is clear that in the time interval $t$ to $t+\tau$ the quantum
representation of the radioactivity source and the live cat does not involve
a radioactivity source that has decayed and a dead cat.

At $t+\tau$, however, if radiation is emitted from the source and
precipitates the poisoning and death of the cat, then we have
an entirely new situation, that is, 
two new systems, two new entities each of which is separable from,
and uncorrelated with its environment.  The source is a system with
fewer radioactive nuclei than were present at time $t$, and the cat is
an entirely new system because a dead cat 
does not need to and does not interact with any 
life support
systems, that is, the interaction between the radioactivity
source
and the live cat induces radical changes in both systems.
At this time, the Hilbert space for the two systems
is $\mathcal{H}_{\r2}\otimes\mathcal{H}_{{\rm dc}}$, and the probability
catalog is 
$\mbox{\Pifont{psy}r}_{\r2}
\otimes 
\mbox{\Pifont{psy}r}_{{\rm dc}}$, 
where the subscript r2 denotes
the radioactivity source with fewer radioactive nuclei than at time
$t$, and dc the dead cat.  It is clear that both the new systems
and the new probability catalogs have no part that refers to the source
at time $t$ and a live cat.   Said differently, no inference
can be made about the radioactivity source prior to decay and the
live cat  by studying the radioactivity source after the decay
and the dead cat because 
in each of the two intervals $t$ to $t + \tau$ and 
(time) $> t + \tau$ 
each of the two entities under consideration is both separable 
from and
uncorrelated with its environment, that is, can be identified
as a system.

After a discussion of theories of measurement, Schr\"odinger
considers two systems that interact with each other for a certain
time and then are separated.  He says ...

\begin{quote}
\hspace{.25truein}``This is the point.  
Whenever one has a complete expectation-catalog 
-- a maximum total knowledge -- a $\psi$-function  for two
completely
separated bodies, or, in better terms, for each of them singly, then
one obviously has it also for the two bodies together, i.e., if one
imagines that neither of them singly but rather the two of them
together make up the object of interest, of our questions about
the future.''

\hspace{.25truein}``But the converse is not true.  {\it Maximal knowledge of a
total system does not necessarily include total knowledge of all
its parts, not even when these are fully separated from each other
and at the moment are not influencing each other at all.}
Thus it may be that some part of what one knows may pertain to
relations or stipulations between the two subsystems (we shall
limit ourselves to two), as follows: if a particular measurement
on the first system yields {\it this} result, then for a particular
measurement on the second the valid expectation statistics are
such and such; but if the measurement in question on the first system
should have {\it that} result, then some other expectation holds for that on
the
second; should a third result occur for the first, then still another
expectation applies to the second; and so on, in the manner of a
complete
disjunction of all possible measurement results which the one
specifically contemplated measurement on the first system can yield.
In this way, any measurement process at all or, what amounts to the
same, any variable at all of the second system can be tied to the
not-yet-known value of any variable at all of the first, and of course
{\it vice versa} also.  If that is the case, if such
conditional statements occur in the combined catalog, 
{\it then it can not possibly be the maximal in regard to the
individual systems.}  For the content of two maximal
individual catalogs would by itself suffice for a maximal 
combined catalog; the conditional statements could not be
added on.''

\hspace{.25truein}``{\it The insufficiency of the $\psi$-function as model 
replacement rests solely on the fact that one doesn't always have
it.}  If one does have it, then by all means let it serve
as description of the state.  But sometimes one does not have it,
in cases where one might reasonably expect to.  And in that case,
one dare not postulate that it ``is actually a particular one, one
just doesn't know it''; the above-chosen standpoint forbids this.
``It'' is namely a sum of knowledge; and knowledge that no ones knows, 
is none.''

\hspace{.25truein}``We continue.  
That a portion of the knowledge should float in
the form of disjunctive conditional statements {\it between} 
the two systems can certainly not happen if we bring up the
two from opposite ends of the world and juxtapose them without
interaction.  For then indeed the two ``know'' nothing about
each other.  A measurement on one cannot possibly furnish
any grasp of what is to be expected of the other.  
Any ``entanglement of predictions'' that takes place can
obviously only go back to the fact that the two bodies
at some earlier time formed in a true sense {\it one} system,
that is were interacting, and have left behind {\it traces}
on each other.  If two separated bodies, each by itself known
maximally, enter a situation in which they influence each other,
and separate again, then there occurs regularly that which I have
just called ``entanglement'' of our knowledge of the two
bodies.  The combined expectation-catalog consists initially 
of a logical sum of the individual catalogs; during the process
it develops causally in accord with known law (there is no
question whatever of measurement here).  The knowledge remains
maximal, but at its end, if the two bodies have again separated,
it is not again split into a logical sum of knowledges about the
individual bodies.  What still remains of {\it that} may 
have  become
less than maximal, even very strongly so. -- One notes the 
great difference over against the classical model theory, where
of course from known initial states and with known interaction
the individual end states would be exactly known.''
\end{quote}

Earlier we discuss most of the points raised in the
preceding statements except for entanglement.  In our terminology,
entanglement means that no two systems can be defined after
the interactions because the two parts may be separable 
from but
correlated with each other (see Section 3.2.1 and A I).

We return to the issue of entanglement in Section 4 where we find
that it is misinterpreted and misapplied.

\subsection{Probability relations between separated systems}

The title of this section is identical to the title of both
communications 
that were presented on Schr\"odinger's behalf at the Cambridge
Philosophical Society by Born [5] and Dirac [6].  In both
communications Schr\"odinger provides detailed mathematical
relations that presumably describe what happens to two systems
both before and after a temporary interaction.

\subsubsection{Born communication}

\hspace{.25truein}In the 
communication presented by Born [5] Schr\"odinger says:

\begin{quote}
\hspace{.25truein}``When two systems, of which we know the states by their
respective representatives, enter into temporary physical 
interaction due to known forces between them, and when
after a time of mutual influence the systems separate again, then
they can no longer be described in the same way as before, viz. 
by endowing each of them with a representative of its own.
I would not call that {\it one} but rather {\it the}
characteristic trait of quantum mechanics, the one that enforces
its entire 
departure from classical lines of thought.  By the interaction
the two representatives
(or $\psi$-functions) have become
entangled.  To disentangle them we must gather further 
information by experiment, although we knew as much as anybody
could possibly know about all that happened.  Of either system,
taken separately, all previous knowledge may be entirely lost,
leaving us but one privilege; to restrict the experiments to
one only of the two systems.  After re-establishing one 
representative by observation, the other one can be inferred
simultaneously.  In what follows the whole of this procedure
will be called {\it the disentanglement}.  Its sinister importance
is due to its being involved in every measuring process and therefore
forming the basis of the quantum theory of measurement,
threatening us thereby with at least a {\it regressus in infinitum},
since it will be noticed that the procedure itself involves
measurement.''
\end{quote}

As we discuss earlier, in quantum theory the definition
of a system requires that it be separable from 
and uncorrelated with
its environment (see A I).  In principle, separability and lack
of correlations are subject to experimental verification.  For
example, if the probabilities of the whole are found to be
described by $\psi(\x_1,\x_2)$, and the probabilities of
the two parts by $\phi_1(\x_1)$ and $\phi_2(\x_2)$,
respectively, then two systems are identifiable if and only if 
\begin{displaymath}
\psi (\x_1,\x_2) = \phi_1(\x_1)\phi_2(\x_2) \nonumber
\end{displaymath}
On the other hand, if the experimental results are such that
\begin{displaymath}
\psi (\x_1,\x_2) \not= \phi_1(\x_1)\phi_2(\x_2) \nonumber
\end{displaymath}
then no two systems can be identified and, therefore, the
two parts may be separable but entangled or correlated.
Separability depends on the lack of forces between the constituents
of the two systems, whereas lack of correlations depends on the
lack of joint 
quantum probabilities between the
constituents of the two systems.

Next, Schr\"odinger states:

\begin{quote}
\hspace{.25truein}``Attention  has recently been called [2]
to the obvious but very
disconcerting fact that even though we restrict the disentangling
measurements to {\it one} system, the representative obtained
for the {\it other} system is by no means independent of the
particular choice of observations which we select for that purpose
and which by the way are {\it entirely} arbitrary.  It is rather
discomforting that the theory should allow a system to be steered or
piloted into one or the other type of state at the experimenter's
mercy in spite of his having no access to it.  This paper does not
aim at a solution of the paradox, it rather adds to it, if possible.
A hint as regards that presumed obstacle will be found at the end.''
\end{quote}

In Sections 2.2 and 3.1 we show that there is neither an EPR nor a cat
paradox.
In particular, the cat paradox is disproven because it is based on
the misconception that a probability catalog -- projector or density 
operator -- can be represented by an expansion in terms of 
(superposition of ?)
two
probability catalogs, each of which applies at a different instant
in time, for example live cat and dead cat.  Such a misconception
is contrary to the structure of all non-relativistic paradigms of
physics because in each of these paradigms
the definitions of the concepts system, property, and state refer to 
one instant in time, and the evolution in time is accounted by the
equation of motion of each paradigm.

The hint alluded to earlier states:

\begin{quote}
``The paradox would be shaken, though, if an observation did not
{\it relate} to a definite moment.  But this would make the present
interpretation of quantum mechanics meaningless, because at present
the {\it objects} of its predictions are considered to be the results
of measurements for definite moments of time.''
\end{quote}

The resolution suggested by the hint is not only contrary to
the meaning of quantum measurements but more importantly it is
detrimental to the most powerful characteristic of modern physics,
namely, the equation of motion of each paradigm.  The equation of
motion allows scientists and engineers to make predictions, and,
as such, builds into each paradigm the opportunity 
for its readjustment
in case the predictions are not consistent with all future
momentary perceptions.  So the idea of considering more than one
instant in time deprives physics of this wonderful
flexibility of readjustment so as to include newer perceptions,
and
thus would  tie each paradigm to
preconceived faith rather than to  experimental 
observations.

\subsubsection{Dirac communication}

\hspace{.25truein}In the introductory remarks of the Dirac communication [6],
Schr\"odinger says:

\begin{quote}
\hspace{.25truein}``1. An earlier paper [5] 
dealt with the following fact.  If for
a system which consists of two entirely separated systems the
representative (or wave function) is known, then the current
interpretation of quantum mechanics obliges us to admit
{\it not only} that by suitable measurements, taken on {\it one}
of the two parts only, the state (or representative or wave
function) of the {\it other} part can be determined without
interfering with it, {\it but also} that, in spite of this
non-interference, the state arrived at {\it depends} quite
decidedly on {\it what} measurements one chooses to take 
-- not only on the results they yield.  So the experimenter, even
with this indirect method, which avoids touching the system itself,
controls its future state in very much the same way as is
well known in the case of a direct measurement.  In this paper
it will be shown that the control, with the indirect measurement, 
is in general not only {\it as} complete but even more complete.
For it will be shown that {\it in general} a sophisticated
experimenter
can, by a suitable device which does {\it not} involve 
measuring non-commuting variables, produce a non-vanishing
probability of driving the system into any state he chooses;
whereas with the ordinary direct method at least the states
orthogonal to the original one are excluded.''

\hspace{.25truein}``The statement is hardly more than a 
corollary to a theorem about ``mixtures'' [7] for which I
claim no priority but the permission of deducing it in the
following section, for it is certainly not {\it well} known.''

\hspace{.25truein}``2. Supposing we knew that a system at a given moment were 
in either one or other of the sequence of states corresponding
to the following sequence of wave functions $\psi_i$, finite or
infinite in number, normalized but in general not orthogonal;
supposing further we knew the probabilities of the system being
in any one of these states -- they must, of course, sum up to
unity and shall be the real positive numbers $w_i$ written in
the second line below the symbol of the function or state:

\[
\left. 
\begin{array}{ccccc}
 \psi_1 & \psi_2 & \psi_3 & \psi_4 & ... \\
 w_1 & w_2 & w_3 & w_4 & ...  \\
 (c^\ast_{1k}) & (c^\ast_{2k}) & (c^\ast_{3k}) & (c^\ast_{4k}) & ...  
\end{array}
\right\} . 
\hspace{2.0truein}  (1) 
\]

In the third line I have written the $k$th development coefficient
of the function above, with respect to an arbitrary 
complete orthogonal system, chosen as a frame of reference
to start with.  The brackets are to indicate that the $k$th
is meant as a representative of all of them; $(c^\ast_{3k})$,
for instance, means all the development coefficients of
$\phi_3$.''

\hspace{.25truein}``The mean value (or expectation value) of a physical variable
$A$, represented in our frame of reference by the matrix
$A_{lk}$, is
\begin{eqnarray}
\bar A & = & w_1 \sum_{l,k} c^{\prime\ast}_{1l} A_{1k}c^\prime_{1k} + w_2
\sum_{l,k} c^{\prime\ast}_{2l} A_{lk}c^\prime_{2k} + ... \nonumber \\
& = & \sum_{l,k} A_{lk} \sum_i w_i c^\prime_{ik} c^{\prime\ast}_{il}. 
\nonumber
\end{eqnarray}
Let us call $U$ the object, represented by the matrix whose
$(k,l)$th element is
\begin{displaymath}
\hspace{2.0in}
U_{kl} = \sum_i w_i c^\prime_{ik} c^{\prime\ast}_{il} \ ,
\hspace{2.0in} (2) \\
\end{displaymath}
then
\begin{displaymath}
\hspace{2.0in} 
\ \bar{A} =\  {\rm Trace}\  (AU).
\hspace{2.0in} (3)
\end{displaymath}
This trace (i.e. sum of diagonal terms) is obviously independent
of the frame of reference, since the $w_i$ are, by their meaning,
invariants and $U$ therefore transforms like a matrix representing
a physical variable, e.g., like $A$.''

\hspace{.25truein}``Since the mean values are all that quantum mechanics predicts
at all, the knowledge of $U$ in a definite frame of reference
{\it exhausts} our real knowledge of the situation, just
as in the case of a ``state'' the wave function exhausts it.
The detailed times (1) from which $U_{kl}$ is composed may
refer to the origin of our knowledge.  But if another set of
similar times leads to the same $U$, then it would be entirely
meaningless to distinguish between the two physical
situations.  $U$ is von Neumann's Statistical Operator.
Its matrix is hermitian.  It has the formal character of a 
real physical variable, but the physical meaning of a wave
function, that is to say it describes the instantaneous
physical situation of the system.''

\hspace{.25truein}''We propose to find {\it all} the different ways (or detailed
data like (1)) which lead to the {\it same U}.  Mark first,
that the hermitian $U_{kl}$ is composed linearly, with positive
coefficients $w_i$, whose sum total is unity, from matrices
each of which obviously has the eigenvalues 1, 0, 0, 0, ..., and
therefore the trace 1.  It follows that $U_{kl}$ has itself the
trace 1 and non-negative eigenvalues.  Now let us change the frame
of reference by a canonical transformation, which makes $U$
diagonal.  Let
\begin{displaymath}
\hspace{2.0in}
p_1, p_2, p_3, p_4, ... \hspace{2.0in} (4)
\end{displaymath}
be this diagonal, the eigenvalues of $U$; according to what has
just been said, they are non-negative and their sum is 1.  This shows
that the {\it same} mixture is obtained by mixing
{\it the orthogonal functions} (or states) which form the basis
of the new frame of reference, in the proportions (or with
the probabilities) $p_k$.''
\end{quote}

Earlier we review the question whether an experimenter,
sophisticated or not, can determine the wave function of a system
without performing direct measurements on it, and find a negative
answer.  We return to the issue here because of the association 
that Schr\"odinger makes 
between expectation values and 
a density operator $U$, defined
as a statistical average of wave functions or projectors.
To be
sure, this is contrary to our proven result that density operators
in quantum theory are defined exclusively by quantum mechanical
probabilities, and not by quantum mechanical probabilities
represented by projectors, and statistical probabilities of
statistical
theories of physics.
However, Schr\"odinger's
discussion of $U$ 
contradicts his definition, and makes $U$ a density operator
that is irreducible or unambiguous (see A V).  Specifically,
he emphasizes that ``$U$ exhausts our real knowledge of
the situation'' and therefore that ``there is no way of 
partitioning $U$ into a statistical average of projectors -- wave
functions''.  Moreover, he asserts that ``$U$ is determined
by the only measurable quantities, that is the expectation
values.''  All these remarks are consistent with the exposition
of quantum theory in the Appendix, and not with von Neumann's
statistical interpretation of density operators (see
A XIV).

\subsection{Summary of the comments on Schr\"odinger's response
to the EPR paper}

\hspace{.25truein}Our summary about Schr\"odinger's responses is to a large
extend similar to the comments we made about the EPR paper.
Three issues that need to be singled out here are the
missed opportunity to recognize the importance of nonstatistical
density operators, the suggested de-emphasis of the equation
of motion as a predictor of events that have not yet been
observed, and the incomplete characterization of entanglement.

\section{Quantum Entanglement: A Modern Perspective}

\subsection{General Remarks}

\hspace{.25truein}The title of this section is identical to the title
of an article that appeared in Physics Today [19].  
We discuss it here because it reflects several misconceptions
that are present in many of the publications on quantum 
computation and quantum information.  The authors of the
article begin with 
brief discussions of the EPR paper and Schr\"odinger's
cat paradox, illustrated by the cartoon shown in Figure 1,
and state;

\begin{quote}
\hspace{.25truein}``Erwin Schr\"odinger
coined the word {\it entanglement} in 1935 in a three-part paper [3]
on the ``present situation in quantum mechanics.''  His article was
prompted by Albert Einstein, Boris Podolsky, and Nathan Rosen's
now celebrated EPR paper that had raised fundamental questions
about quantum mechanics earlier that year. ...
For the three decades following the 1935 articles, the debate about
entanglement and the ``EPR dilemma'' -- how to make sense of the
presumably nonlocal effect one particle's measurement has on
another -- was philosophical in nature, and for many physicists it
was nothing more than that.  The 1964 publication by 
John Bell [20] changed
that situation dramatically.  Bell derived correlation inequalities
that can be violated in quantum mechanics but have to be satisfied
within every model that is local and complete --- so-called
local hidden-variable models.  Bell's work made it possible to
test whether local hidden-variable models can account for observed
physical phenomena.  Early and ongoing recent experiments showing
violations of such Bell inequalities have invalidated local
hidden-variable models and lend support to the quantum-mechanical
view of nature.  In particular, an observed violation of a Bell
inequality demonstrates the presence of entanglement in
a quantum system.''
\end{quote}

Earlier we show that there are no EPR and Schr\"odinger's
cat paradoxes.  In addition 
we show that entanglement cannot occur in a system because,
by definition, a system must be both separable from and 
uncorrelated with its environment (see A I).  If both these
conditions are not satisfied then no system can be defined.
Moreover, entanglement is not related to hidden variables.

\subsection{Entanglement for the 21st century}

\hspace{.25truein}The authors of Ref. [19] say:

\begin{quote}
\hspace{.25truein}``An experimentalist, Alice, wishes to send an unknown
state $| s \big> = \alpha|0 \big> + \beta|1 \big>$ of a two level
quantum system to another experimentalist, Bob, in a distant
laboratory. ... Alice and Bob do not have the means of
directly transmitting the quantum system from one place to
another ... but let us imagine that they do share an entangled
state.  Consider the  case in which Alice and Bob each
have one spin of a shared singlet state of two 
spin-1/2 particles \linebreak
 $|\Psi^- \big> = 1/\sqrt{2}(|\uparrow,\downarrow\big> 
+ |\downarrow,\uparrow\big>)$, also called EPR pair.  Alice can transmit
her spin $|s\big>$ to Bob by performing a certain joint measurement
on her spin $|s\big>$ and her half of the EPR pair. She tells Bob
the result of her measurement and depending on her information,
Bob rotates his half of the EPR pair to obtain the state
$|s\big>$.''
\end{quote}

There are three statements in the preceding quote that we find
contrary to quantum theory: 
(i) In principle, an infinite number of
measurements on an ensemble 
of identical systems, identically prepared
is needed to determine
$|s\big>$, regardless of the values of $\alpha$ and $\beta$
$(\alpha^2+\beta^2=1)$.
So how can one measurement reveal any information about
$|s\big>$?;  (ii) The ket $|\Psi^-\big>$ is not
an EPR state.  It is an expansion of $|\Psi^-\big>$
({\it not a superposition}) in terms of two orthonormal eigenkets
of a two-spin system; and (iii) a system is not
a state, and a state is not a system.

Next, the authors of Ref. [19] say:

\begin{quote}
\hspace{.25truein}``The spin-singlet EPR state that Alice and Bob share in
quantum teleportation is called a maximally entangled state.
Even though the two spins together constitute a definite
pure state, each spin state is maximally undetermined or mixed
when considered separately.  
In mathematical terms, Alice's local density 
matrix --- obtained by tracing over Bob's spin degrees of freedom,
${\rm Tr}_B(|\Psi^-\big> \big<\Psi^-|)$ --- has 
equal probability for spin up and
spin down.
In keeping with Schr\"odinger's understanding of entanglement,
one measures the amount of entanglement in a general pure
state $\phi$ in terms of the lack of information about its
local parts.  The von Neumann entropy 
$S(\mbox{\Pifont{psy}r})
= -{\rm Tr}(
\mbox{\Pifont{psy}r}
{\rm log}
\mbox{\Pifont{psy}r})$
is used as a measure of that information.  In other words, the
entropy of entanglement $E$ of the pure state $\phi$ is equal
to the von Neumann entropy of, say, Alice's density matrix
$\mbox{\Pifont{psy}r}
= {\rm Tr}_B|\phi\big>\big<\phi|$.''
\end{quote}

The statements just cited misrepresent the theory of quantum
phenomena.  The trace of any projector either
$|\Psi^-\big>\big<\Psi^-|$ or $|\phi \big>\big<\phi|$ is equal to unity
as can be readily and easily shown by finding the density
matrix in a representation in a Hilbert space that has
either $|\Psi^-\big>\big<\Psi^-|$ or $|\phi\big>\big<\phi|$ 
as one of its mutually orthogonal dimensions.  Accordingly, the
{\it thermodynamic entropy} of any projector is equal to zero.
The von Neumann entropy is not relevant to this discussion, 
let alone the fact that it does not represent the entropy of
thermodynamics (see A V).

\subsection{Superpositions of versus expansions in terms
of either kets or projectors}

\hspace{.25truein}In many discussions of either quantum mechanics or quantum
computation and quantum information, expressions of the forms
\begin{displaymath}
| \Psi \big> = \sum_\i a_\i|u_\i \big> ~~ {\rm or} ~~ 
\mbox{\Pifont{psy}r}
= \sum_\i {\rm p}_\i | u_\i \big> \big< u_\i | 
\end{displaymath}
are interpreted as superpositions of $|u_\i\big>$'s or
$|u_\i\big>\big<u_\i|$'s, respectively.  For example, the ket
$(|0\big>+|1\big>)/\sqrt{2}$ is pictorially represented as shown
in Figure 2 reproduced from Ref.~[1].  
Such a representation, however, 
is not correct because the ket just cited is not a 
superposition of two electronic levels 
or two spins 
of an atom but
an expansion 
of the probability catalog of one spin 
in terms of two orthonormal kets of one spin.  In
general, any ket can be represented by an 
expansion in terms of a 
complete set of orthonormal eigenkets of
an operator representing an observable,
and of course, different expansions result for different
orthonormal eigenkets.  
Similar
remarks can be made about a density operator that is
represented by a homogeneous or irreducible ensemble 
and various expansions in terms of 
complete sets of orthonormal
eigenprojectors of various observables.

\begin{figure}[here]
\epsfxsize=3 truein
\vbox{
\smallskip\centerline{\epsfbox{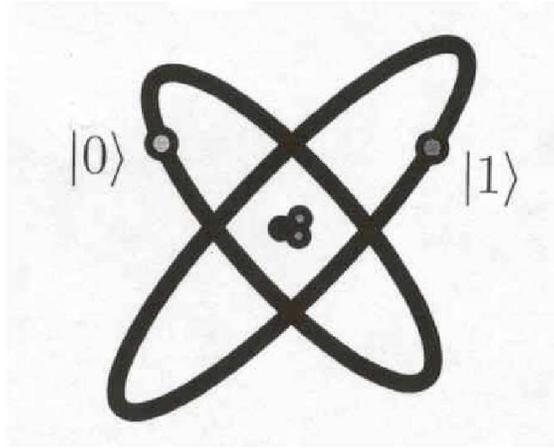}}
\caption{Qubit $(|0>+|1>)/\sqrt{2}$ represented by
two electronic levels of an atom[1].    }}
\end{figure}

Nevertheless, 
the idea of superposition 
instead of expansion 
is so pervasive as to be
considered a special feature of quantum theory.
For example,
Jozsa and Linden state~[21]:

\begin{quote}
``One fundamental non-classical feature of the quantum formalism
is the rule for how the space $S_{AB}$ of a composite
system $AB$ is constructed from the state spaces $S_A$ and $S_B$ 
of the parts $A$ and $B$.  In classical theory, $S_{AB}$ is the
cartesian product of $S_A$ and $S_B$ whereas in quantum
theory it is the tensor product (and the state spaces are
linear spaces).  This essential distinction between cartesian
and tensor products is precisely the phenomenon of quantum
entanglement viz. the existence of pure states of a 
composite system that are not product states of the parts.''
\end{quote}

The statement just cited 
is not correct.  First, the space in question is
not the state space but the Hilbert space in which the
probability catalogs (kets and/or density operators) are
depicted.  Second, as we discuss earlier, the tensor product
is valid if and only if two systems can be identified, that is,
if and only if two entities can be specified each of which is
separable from and uncorrelated with its environment.  If both
requirements are not satisfied by the parts but are satisfied
by their composite, then the composite is identified as a system,
and entanglement has no meaning.  Similar concerns about the
meaning of the concept of entanglement are expressed by Wojcik
[22].

Next, Jozsa and Linden state

\begin{quote}
``In quantum theory, state spaces always admit the superposition
principle whereas in classical theory, state spaces generally do
not have a linear structure.  But there do exist classical state
spaces with natural linear structure (e.g., the space of states
of an elastic vibrating string with fixed end points) so the
possibility of superposition {\it in itself}, is not a uniquely
quantum feature.''
\end{quote}

As we discuss earlier, this statement is not correct.
Expansions of either kets or density operators are not
superpositions, and cannot be visualized as ``elastic vibrating
strings with fixed endpoints.''  In general, we feel 
very strongly that
greater attention should be given to the use or abuse of the
concept of entanglement because its correct consideration may
disclose new approaches to the challenge of development
of quantum computation and quantum information. 

\section{Conclusions}

\hspace{.25truein}Upon detailed and close scrutiny of the ideas presented in
the EPR paper and in Schr\"odinger's responses to this paper,
we find many faulty conclusions about the very successful
theories of quantum mechanics and thermodynamics.
We show that the faulty conclusions are due to lack of correct
definitions of basic concepts, use of a postulate that is
proven not to be valid, and misinterpretations of key theorems
of the theory.

We also find that the faulty conclusions have permeated the
theoretical underpinnings of quantum computation and quantum
information.
Our criteria are based on an exposition of quantum theory
(see Appendix) that unifies quantum and thermodynamic ideas
without resort to statistical probabilities.

We hope our review of the paradoxologies presented in both the
EPR paper and in Schr\"odinger's responses will be helpful
in the development of a more rigorous approach to the
fascinating and promising fields of quantum computation and
quantum information.

\section*{Appendix: Quantum Theory}
\setcounter{equation}{0}
\renewcommand{\theequation}{A-\arabic{equation}}
\setcounter{figure}{0}
\renewcommand{\thefigure}{A-\arabic{figure}}

\hspace{.25truein}In this Appendix we present a summary of nonrelativistic quantum
theory that differs from the presentations in most textbooks
on the subject.  The key differences are the discoveries
that for a broad class of quantum-mechanical problems:
(i) the probabilities associated with ensembles of measurement
results at an instant in time require a mathematical concept
delimited by but more general than a wave function or projector;
and (ii) the evolution in time of the new mathematical concept
requires  an equation of motion delimited by but more general
than the Schr\"odinger equation.
Our definitions, postulates, and major theorems of quantum
theory are based on slightly modified statements made
by Park and Margenau [23], and close scrutiny of the
implications of these statements.

The first difference is the recognition by Hatsopoulos and
Gyftopoulos [8] that there exist two classes of quantum
problems.  In the first class, the quantum-mechanical
probabilities associated with measurement results are fully
described by wave functions or projectors, whereas in the second
class the probabilities just cited require density operators that
involve no statistical averaging over projectors --- no
mixtures of quantum and statistical probabilities.  The
same result emerges from the excellent review of the foundations
of quantum mechanics by Jauch [9].

The second difference is the recognition that the evolution
in time of non-statistical density operators requires a nonlinear
equation of motion, and the discovery of such an equation by
Beretta et al [24,25].

\subsection*{Kinematics: Definitions, postulates, and theorems
 at an instant in time.}

A I: {\it System}.  The term {\it system} means a set of
specified types and amounts of constituents, confined and
controlled by a nest of internal and external forces.
For example, one hydrogen atom consisting of an electron
and a proton confined in a three-dimensional 
cubical potential
well of 
side size a.
The internal force arises from the Coulomb interaction between
the proton and the electron, and depends on the spatial coordinates
of both constituents.  The external force arises from the
impermeable walls of the cubical box, that is, a force equal
to infinity at every point on the inside surface of
the cube.
This force 
depends only on the spatial
coordinates of either the electron or the proton or the hydrogen
atom as a unit and not on any coordinates of constituents that
are not included in the system, that is, the system is
{\it separable} from its environment.  In addition, in order
to be totally independent  and fully identifiable, the system must
also be {\it statistically uncorrelated} with its environment.
In general, this system definition is valid for any
paradigm of physics, and sine qua non for quantum theory.

A II: {\it System postulate.}   To every system there corresponds
a complex, separate, complete, inner product space, a Hilbert
space $\mathcal{H}$.  The Hilbert space of a composite system of
two distinguishable and identifiable subsystems 1 and 2, with
associated
Hilbert spaces $\mathcal{H}^1$ and $\mathcal{H}^2$, respectively,
is the direct product space $\mathcal{H}^1 \otimes \mathcal{H}^2$.

A III: {\it Homogeneous or unambiguous ensemble}.  At an instant in
time, an ensemble of identical systems is called
{\it homogeneous} or {\it unambiguous} if and 
only if upon subdivision into subensembles in any conceivable way that
does not perturb any member, each subensemble yields in every
respect measurement results -- spectra of values and frequency of
occurrence of each value within a spectrum -- identical to the
corresponding results obtained from the ensemble.  For example, 
the spectrum of energy measurement results and the frequency
of occurrence of each energy measurement result obtained from any
subensemble are identical to the spectrum of energy measurement
results and the frequency of occurrence of each energy measurement
result obtained from an independent ensemble that includes all
the subensembles.  Other criteria are presented in Ref.~[8].

A IV: {\it Preparation}.  A {\it preparation} is a reproducible scheme
used to generate one or more 
homogeneous ensembles for study.

A V: Every textbook on quantum mechanics avers that the probabilities
associated with measurement
results\footnote{We 
use the expression
``probabilities associated with measurement results'' rather than
``state'' because the definition of state (see A XIV) requires more
than the specification of a projector or a density operator.}
 of a system in a state ``i'' are
described by a wave function $\Psi_\i$
or, equivalently, a projector 
$|\Psi_\i \big>\big< \Psi_\i| = 
\mbox{\Pifont{psy}r}_\i = \mbox{\Pifont{psy}r}_\i^2$ and that
each density operator
$\mbox{\Pifont{psy}r} > \mbox{\Pifont{psy}r}^2$ 
is a  statistical average of projectors, that is, each
$\mbox{\Pifont{psy}r}$ represents a mixture of
quantum mechanical probabilities determined by projectors and
nonquantum-mechanical or statistical probabilities that reflect our
inability to make difficult calculations, our lack of interest in
details, and our lack of knowledge of initial conditions.  Mixtures
have been introduced by von Neumann [7] for the purpose of explaining
thermodynamic equilibrium phenomena in terms of statistical quantum
mechanics (see also Jaynes [26] and Katz [27]).

Pictorially, we can visualize a projector by an ensemble of identical
systems, identically prepared.  Each member of such an ensemble is
characterized by the same projector
$\mbox{\Pifont{psy}r}_\i$, and von Neumann calls the ensemble
{\it homogeneous}.  Similarly, we can visualize a density operator
$\mbox{\Pifont{psy}r}$ 
consisting of a statistical mixture of two projectors
$\mbox{\Pifont{psy}r}=\alpha_1\mbox{\Pifont{psy}r}_1+
\alpha_2\mbox{\Pifont{psy}r}_2, \alpha_1+\alpha_2=1,  
\mbox{\Pifont{psy}r}_1 \not= \mbox{\Pifont{psy}r}_2 \not=
\mbox{\Pifont{psy}r}$, 
where 
$\mbox{\Pifont{psy}r}_1$ and
$\mbox{\Pifont{psy}r}_2$ represent quantum-mechanical probabilities,
$\alpha_1$ and $\alpha_2$ represent 
statistical probabilities, and
the ensemble is called {\it heterogeneous} or
{\it ambiguous} [8].  A pictorial representation of a heterogeneous
ensemble is shown in Figure A-1.

These results beg the questions (i) Are there quantum-mechanical
problems that involve probability distributions which cannot be
described by a projector but require a purely quantum-mechanical
density operator --- a density operator which is not a statistical
mixture of projectors?; and (ii) Are such purely
quantum-mechanical
density operators consistent with the foundations of
quantum mechanics?

\begin{figure}
\begin{center} 
\begin{tabular}{cccccc} 
\multicolumn{6}{c}{{\sf HETEROGENEOUS ENSEMBLE}}\\[2ex] 
{\makebox[.3in][c]{\framebox{$\mbox{\Pifont{psy}r}_1$}}} & 
   {\makebox[.3in][c]{\framebox{$\mbox{\Pifont{psy}r}_1$} }} & 
   {\makebox[.3in][c]{\framebox{$\mbox{\Pifont{psy}r}_1$} }} & 
   {\makebox[.3in][c]{$\cdots$ }} & 
   {\makebox[.3in][c]{\framebox{$\mbox{\Pifont{psy}r}_2$}}} & 
   {\makebox[.3in][c]{\framebox{$\mbox{\Pifont{psy}r}_2$}}}\\ 
& & & $ \mbox{\Pifont{psy}r} \neq \mbox{\Pifont{psy}r}_1$ & & \\[1ex] 
{\makebox[.3in][c]{\framebox{$\mbox{\Pifont{psy}r}_1$}}} & 
   {\makebox[.3in][c]{\framebox{$\mbox{\Pifont{psy}r}_1$} }} & 
   {\makebox[.3in][c]{\framebox{$\mbox{\Pifont{psy}r}_1$} }} & 
   {\makebox[.3in][c]{$\cdots$ }} & 
   {\makebox[.3in][c]{\framebox{$\mbox{\Pifont{psy}r}_2$}}} & 
   {\makebox[.3in][c]{\framebox{$\mbox{\Pifont{psy}r}_2$}}}\\ 
& & & $ \mbox{\Pifont{psy}r} \neq \mbox{\Pifont{psy}r}_2$ & & \\[1ex] 
{\makebox[.3in][c]{\framebox{$\mbox{\Pifont{psy}r}_1$}}} & 
   {\makebox[.3in][c]{\framebox{$\mbox{\Pifont{psy}r}_1$} }} & 
   {\makebox[.3in][c]{\framebox{$\mbox{\Pifont{psy}r}_1$} }} & 
   {\makebox[.3in][c]{$\cdots$ }} & 
   {\makebox[.3in][c]{\framebox{$\mbox{\Pifont{psy}r}_2$}}} & 
   {\makebox[.3in][c]{\framebox{$\mbox{\Pifont{psy}r}_2$}}}\\[2ex] 
$\vdots$ &  $\vdots$ & $\vdots$ & $\cdots$ & $\vdots$ &$\vdots$ \\[1.5ex] 
\multicolumn{3}{c}{ $ 
     {\mbox{\footnotesize\sf FRACTION}} \  \mbox{\Pifont{psy}a}_1 $ } & 
   \multicolumn{1}{c}{{ \quad }} & 
   \multicolumn{2}{c}{ $ 
     {\mbox{\footnotesize\sf FRACTION}} \  \mbox{\Pifont{psy}a}_2 $ } \\ 
\multicolumn{6}{c}{ $ 
     {\mbox{\footnotesize\sf OVERALL DENSITY}} \ \mbox{\Pifont{psy}r} = 
    \mbox{\Pifont{psy}a}_1 \mbox{\Pifont{psy}r}_1 + \mbox{\Pifont{psy}a}_2 
    \mbox{\Pifont{psy}r}_2$ } 
\end{tabular} 
\end{center} 
\caption{Pictorial 
representation of a heterogeneous ensemble. Each of 
the subensembles for 
$\mbox{\Pifont{psy}r}_1$ and 
$\mbox{\Pifont{psy}r}_2$ represents either a 
projector ($\mbox{\Pifont{psy}r}_{\mathrm{i}} 
= \mbox{\Pifont{psy}r}^2_{\mathrm{i}}$) or a density 
operator ($\mbox{\Pifont{psy}r}_{\mathrm{i}} > 
 \mbox{\Pifont{psy}r}^2_{\mathrm{i}}$), for 
i = 1,2, and $\alpha_1 + \alpha_2 = 1$}
\end{figure} 

\begin{figure}
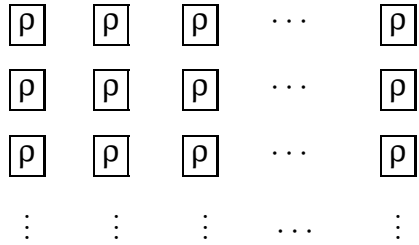
 
\begin{center} 
\begin{tabular}{ccccc} 
\multicolumn{5}{c}{{\sf HOMOGENEOUS ENSEMBLE}}\\[2ex] 
{\makebox[.3in][c]{\framebox{$\mbox{\Pifont{psy}r}$}}} & 
   {\makebox[.3in][c]{\framebox{$\mbox{\Pifont{psy}r}$} }} & 
   {\makebox[.3in][c]{\framebox{$\mbox{\Pifont{psy}r}$} }} & 
   {\makebox[.3in][c]{$\cdots$ }} & 
   {\makebox[.3in][c]{\framebox{$\mbox{\Pifont{psy}r}$}}}\\[2ex] 
{\makebox[.3in][c]{\framebox{$\mbox{\Pifont{psy}r}$}}} & 
   {\makebox[.3in][c]{\framebox{$\mbox{\Pifont{psy}r}$} }} & 
   {\makebox[.3in][c]{\framebox{$\mbox{\Pifont{psy}r}$} }} & 
   {\makebox[.3in][c]{$\cdots$ }} & 
   {\makebox[.3in][c]{\framebox{$\mbox{\Pifont{psy}r}$}}}\\[2ex] 
{\makebox[.3in][c]{\framebox{$\mbox{\Pifont{psy}r}$}}} & 
   {\makebox[.3in][c]{\framebox{$\mbox{\Pifont{psy}r}$} }} & 
   {\makebox[.3in][c]{\framebox{$\mbox{\Pifont{psy}r}$} }} & 
   {\makebox[.3in][c]{$\cdots$ }} & 
   {\makebox[.3in][c]{\framebox{$\mbox{\Pifont{psy}r}$}}}\\[2ex] 
$\vdots$ &  $\vdots$ & $\vdots$ & $\cdots$ & $\vdots$ \\[1.5ex] 
\multicolumn{5}{c}{ $ 
     {\mbox{\footnotesize\sf OVERALL DENSITY OPERATOR}} = 
\mbox{\Pifont{psy}r} $} 
\end{tabular} 
\end{center} 
\caption{Pictorial 
representation of a homogeneous ensemble. Each of the 
members of the ensemble is characterized by the same 
density operator $\mbox{\Pifont{psy}r} \geq \mbox{\Pifont{psy}r}^2$.  
It is clear that any conceivable subensemble of a 
homogeneous ensemble is characterized by 
the same $\mbox{\Pifont{psy}r}$ as the ensemble.} 
\end{figure} 

Upon close scrutiny of the definitions, postulates, and key theorems of
quantum theory, we find that the answers to both questions are yes.
These answers were discovered by Hatsopoulos and Gyftopoulos [8]
in the course of their development of a unified quantum theory of
mechanics
and thermodynamics, and by Jauch [9] in his systematic and
rigorous analyses of the foundations of quantum mechanics
(see also Section 3.2.2).

Pictorially, we can visualize a purely quantum-mechanical
density operator
$\mbox{\Pifont{psy}r} > \mbox{\Pifont{psy}r}^2$
by an ensemble of identical systems, identically prepared.
Each member of such an ensemble is characterized by the same
$\mbox{\Pifont{psy}r}$ as shown in Figure A-2; and by analogy
to the results for a projector, we call this ensemble
{\it homogeneous} or {\it unambiguous} [8].  If the density
operator is a projector 
$\mbox{\Pifont{psy}r}_\i = \mbox{\Pifont{psy}r}_\i^2$, 
then each member of the ensemble is characterized by the same
$\mbox{\Pifont{psy}r}_\i$
as originally proposed by von Neumann.

The recognition of the existence of density operators that correspond
to homogeneous ensembles has many interesting implications.  It
extends quantum ideas to thermodynamics, and thermodynamic principles
to quantum phenomena.

For example, it is shown that thermodynamics applies to all systems
(both large and small, including one particle systems, such as one
spin),
to all states (both thermodynamic equilibrium, and not
thermodynamic equilibrium), and that entropy is 
an intrinsic property
of each constituent of a system [28] (in the same sense that
inertial mass is a property of each constituent) and not a measure
of ignorance, or lack of information, or disorder [29].
Again, it is shown that entropy is a measure of the quantum-mechanical
spatial shapes of the constituents of a system, and that
irreversibility is solely due to the changes of these shapes as
the constituents try to conform to the external and
internal force fields of the system [11,12].  Incidentally, for
a spin, entropy is a measure of the orientation of the spin 
within or on  the
Block sphere (Figure A-3).

\begin{figure}[here]
\epsfxsize=3 truein
\vbox{
\smallskip\centerline{\epsfbox{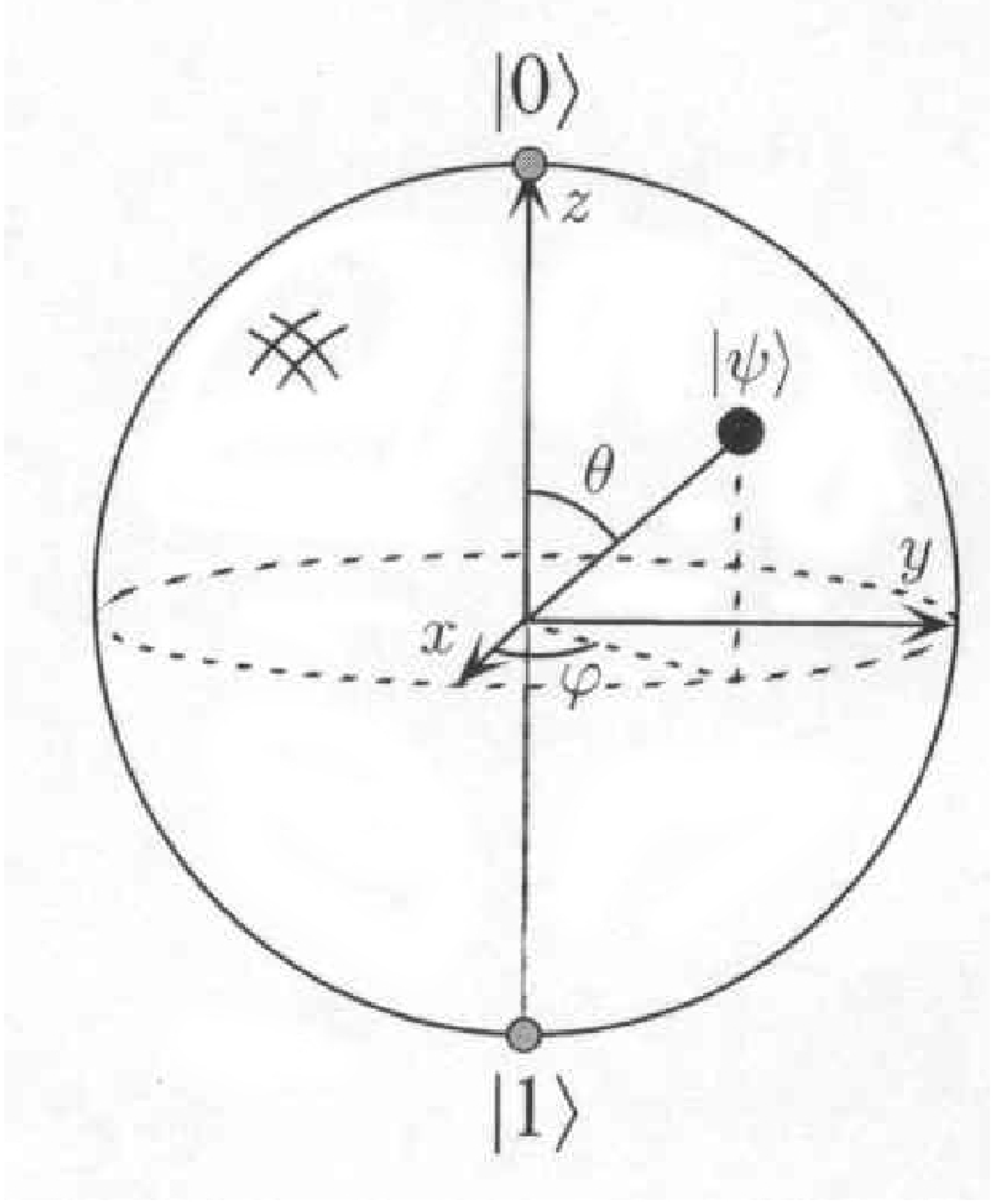}}
\caption{Block sphere representation of a qubit.}}
\end{figure}
\vskip .50truein

It is noteworthy that for given values of energy, volume, and
amounts of constituents, if
$\mbox{\Pifont{psy}r}$ is a projector then the entropy
$S$=0, if $\mbox{\Pifont{psy}r}$ is neither a projector
nor a statistical average of projectors, and
corresponds to a state which is not stable equilibrium
(not thermodynamic equilibrium), then $S$ has a positive
value smaller than the largest possible 
value for the given specifications,
and if $\mbox{\Pifont{psy}r}$ corresponds to the unique
stable equilibrium state, then $S$ has the largest value of
all the entropies of the system which share the given values of
energy, volume, and amounts of constituents.  Said differently,
all projectors or wave functions correspond to zero entropy
physics, all largest entropy density operators for given
system specifications correspond to stable equilibrium states
-- classical thermodynamics -- and all other density operators
that are associated neither with zero entropy, nor with largest
value entropy correspond to probability distributions that can be
represented neither by projectors nor by stable equilibrium state
density operators.
The few results just cited indicate that the restriction of quantum
mechanics to problems that require probability distributions
described only by projectors is both unwarranted and nonproductive.

A VI: {\it Property}.  The term {\it property} refers to any 
attribute of a system that can be quantitatively evaluated at
an instant in time by means of measurements and specified procedures.
All measurement results and procedures are assumed to be precise, that
is, both error free, and unaffected by the
measurement and the measurement procedures.  Moreover, they are
assumed
not to depend on either other systems or other instants in time.

A VII: {\it Observable}.  From the definition just cited, it
follows that each property can be observed.  Traditionally,
however, in quantum theory a property is called an {\it observable}
only if it conforms to the following mathematical representations.

A VIII: {\it Correspondence postulate}.  Some linear
Hermitian operators A, B, ... on Hilbert space $\mathcal{H}$,
which have complete orthonormal sets of eigenvectors, 
correspond to observables of a system.

The inclusion of the word ``some'' in the correspondence
postulate is very important because it indicates that there
exist Hermitian operators that do not represent
observables, and properties that cannot be represented by
Hermitian operators.  The first category accounts for
Wick et al. [30] superselection rules, and the second 
accounts both for
compatibility of simultaneous measurements introduced by Park and
Margenau [23], and for properties, such as temperature, that are not
represented by operators.

A IX: {\it Measurement act.}  A {\it measurement act} is a
reproducible scheme of measurements and operations on a
member of an ensemble.  Regardless of whether the measurement
refers to an observable or not, 
in principle the result   of such an
act is presumed to be precise, that is, 
an error and perturbation free
number.

If a measurement act of an observable is applied to each and every
member of a homogeneous ensemble, the results conform to the following
postulate and theorems.

A X: {\it Mean-value postulate}.  If a measurement act of an 
observable represented by a Hermitian operator A is applied to each
and every member of a homogeneous ensemble, there exists a 
linear functional m(A) of A such that the value of m(A)
equals the arithmetic mean of the ensemble of A measurements,
that is,
\begin{equation}
{\rm m(A)} = \big< {\rm A} \big> = \sum_\i a_\i/{\rm N} \ 
~ {\rm for~ N}  \ \rightarrow \infty  
\end{equation}
where $a_\i$ is the measurement result of a measurement act
of A 
applied to the ith member of the ensemble, a large number 
(theoretically infinite)
of
$a_\i$'s have the same numerical value, and both m(A) and 
$\big<$A$\big>$ 
represent the expectation value of A.

A XI: {\it Mean-value theorem}.  For each of the mean-value
functionals or expectation values m(A) of a system at an instant
in time, there exists the same Hermitian operator 
$\mbox{\Pifont{psy}r}$ such that
\begin{equation}
{\rm m(A) =} \big< {\rm A} \big> = {\rm Tr} [\mbox{\Pifont{psy}r}
{\rm A}]
\end{equation}
The operator $\mbox{\Pifont{psy}r}$ is known as the {\it density
operator} or the {\it density of measurement results of 
observables}, and here it can be represented solely by a homogeneous
ensemble as shown in Figure A-2, that is, each member of the
ensemble is characterized by the same $\mbox{\Pifont{psy}r}$ 
as any other member.
It is noteworthy that the value 
$\big<$A$\big>$ of an observable A depends
exclusively on the Hermitian operator A that represents the
observable and on the density operator $\mbox{\Pifont{psy}r}$,
and not on any other operator that either commutes or does not
commute with operator A.

The operator $\mbox{\Pifont{psy}r}$ is proven to be Hermitian,
positive, unit trace and, in general, not a projector, that is,
\begin{equation}
\mbox{\Pifont{psy}r} > 0; \ {\rm Tr}\mbox{\Pifont{psy}r} = 1;  \ 
{\rm and} \ \mbox{\Pifont{psy}r} \geq \mbox{\Pifont{psy}r}^2
\end{equation}

A XII: {\it Probability theorem}.  If a measurement act of an
observable represented by operator A is applied to each and every
member of a homogeneous ensemble characterized by
$\mbox{\Pifont{psy}r}$,
the probability or frequency of occurrence $W(a_\n)$ that the
results will yield eigenvalue $a_\n$ is given by the relation
\begin{equation}
W(a_\n) = {\rm Tr}[\mbox{\Pifont{psy}r}{\rm A_\n} ]
\end{equation}
where A$_\n$ is the projection onto the subspace belonging
to $a_\n$
\begin{equation}
{\rm A} |\alpha_\n^{(\i)} > = a_\n | \alpha_\n^{(\i)} >
\ \ \  {\rm for \ n} \ = 1, 2, ... \ \ \ 
{\rm and\ i} \ = 1, 2, ..., {\rm g}\ ,
\end{equation}
and g is the degeneracy of $a_\n$.

A XIII: {\it Measurement result theorem}.  The only possible result
of a measurement act of the observable represented by A is
one of the eigenvalues of A (Eq. A-5). 

A XIV: {\it State}.  The term {\it state} means all that can be
said about a system at an instant in time, that is, a set of
Hermitian operators A,B, ... that correspond to a set of
n$^2$-1 linearly independent observables -- the value
of an independent observable can be varied without affecting the
values of other observables -- and the relations

\begin{eqnarray}
\big< {\rm A} \big>  & = & {\rm Tr} [\mbox{\Pifont{psy}r}{\rm A}] =  
\sum_\i a_\i /{\rm N} \nonumber\\
\big< {\rm B} \big> & = & {\rm Tr}[\mbox{\Pifont{psy}r}{\rm B}]  =  
\sum_\i b_\i/{\rm N}  \\                     
\vdots & \vdots & \hspace{.25truein}  \vdots 
\hspace{.25truein}  \vdots \hspace{.25truein}  \vdots  \nonumber 
\end{eqnarray}

\noindent
where n is the dimensionality of the Hilbert space, and N $\rightarrow
\infty$.

In Eqs. (A-6), either the density operator 
$\mbox{\Pifont{psy}r}$ is specified {\it a priori} and the values
of the observables are calculated, or the values $\sum_\i a_\i/{\rm
N}$, $\sum_\i b_\i/{\rm N}, ...$ of the linearly   
independent observables are either specified or, in
principle, experimentally established and a unique operator
$\mbox{\Pifont{psy}r}$ is calculated.  The mappings both from
$\mbox{\Pifont{psy}r}$ to expectation values and from expectation
values
to $\mbox{\Pifont{psy}r}$ 
are unique because Eqs. (A-6) are linear from expectation
values to $\mbox{\Pifont{psy}r}$ and from 
$\mbox{\Pifont{psy}r}$ to expectation values.

It is noteworthy, that no quantum-theoretic requirement
exists which excludes the possibility that the mapping
from the measurable expectation values to 
$\mbox{\Pifont{psy}r}$
must yield a projector $\mbox{\Pifont{psy}r} =
\mbox{\Pifont{psy}r}^2$ rather than a density operator
$\mbox{\Pifont{psy}r} > \mbox{\Pifont{psy}r}^2$, and,
conversely, that a pre-specified operator 
$\mbox{\Pifont{psy}r}$ must necessarily be a projector.
In fact, the existence of density operators that are not
derived as a mixture of quantum probabilities and
statistical probabilities provides the means for the
unification of quantum theory and thermodynamics without
resorting to statistical considerations [8].

It is also noteworthy that only a linear operator X and its
eigenvalues $x_\i$ are included in Eqs. (A-6).  
For example, only the Hamiltonian operator H and its
eigenvalues $\epsilon_1, \epsilon_2, ...,$ appear once
in Eqs. (A-6).  Operators H$^{\rm m}$ and their eigenvalues
$\epsilon_1^{\rm m}, \epsilon_2^{\rm m}, ...$ 
for~m~$>$~1 are excluded.  The reasons for this important 
restriction are: ($i$) For m even, information about the sign
of eigenvalues of any X is lost; ($ii$) $\big< {\rm X} \big>$
and $\big< {\rm X^m}\big>$ for any m $>$ 1 are not linearly
independent; and ($iii$) In general, in the extreme case
that only X$^{\rm m}$
for m = 1, 2, ... are used, then all the off-diagonal 
elements of 
$\mbox{\Pifont{psy}r}$ in the X representation are lost.

Finally, it is clear that the definition of state is not
synonymous either with the concept of a wave function or the
concept of a density operator (A~XI).  
The definition of state 
requires both
the specification of the system (A~I), and either a complete
set of measurable, linearly independent expectation values, 
or a prescribed density operator $\mbox{\Pifont{psy}r} \geq
\mbox{\Pifont{psy}r}^2$ without any statistical probabilities.

A XV: {\it Uncertainty relations.}
Ever since the inception of quantum mechanics, the uncertainty
relation that corresponds to a pair of observables represented
by non-commuting operators is interpreted by many
scientists and engineers as a limitation on the accuracy
with which the observables can be measured.  For example,
Heisenberg [31] writes the uncertainty relation between
position $x$ and momentum $p_x$ in the form
{\renewcommand{\theequation}{A-7}
\begin{equation}
\Delta x \Delta p_x \geq h
\end{equation}
and comments: ``This uncertainty relation specifies the
limits within which the particle picture can be applied.
Any use of the words ``position'' and ``velocity'' 
with an accuracy exceeding that given by equation (A-7) is
just as meaningless as the use of words whose sense is
not defined.''

Again Louisell [32] addresses the issue of uncertainty relations,
and concludes: ... ``That is the actual measurement itself
disturbs the system being measured in an uncontrollable
way, regardless of the care, skill, or ingenuity of the
experimenter. ... In quantum mechanics the precise measurement
of both coordinates and momenta is not possible even in
principle.''

The remarks by Heisenberg and Louisell are representative
of the intrepretations of uncertainty relations discussed in many
textbooks on quantum mechanics [33]. These
remarks, however, cannot be deduced from the postulates and
theorems of quantum theory.

The measurement result theorem avers that the measurement
of an observable is a precise (perturbation free) and,
in many cases, precisely calculable eigenvalue of the operator
that represents the observable.  So neither a measurement 
perturbation nor a measurement 
error is contemplated by the theorem.  An outstanding example
of measurement 
accuracy is the Lamb shift~[34].

The probability theorem avers that we cannot predict which
precise eigenvalue each measurement will yield except in terms
of either a prespecified or a measurable probability or
frequency of occurrence.

It follows that an ensemble of measurements of an observable
performed on an ensemble of identical systems, identically
prepared yields a range of eigenvalues, and a probability
or frequency of occurrence distribution over the
eigenvalues.  In principle, both results are precise and
involve no disturbances induced by the measuring
procedures.

To be sure, each probability distribution of an observable
represented by operator X has a variance
{\renewcommand{\theequation}{A-8}
\begin{equation}
\left( \Delta {\rm X} \right)^2 = {\rm Tr} 
\mbox{\Pifont{psy}r} {\rm X}^2 - \left( {\rm Tr} 
\mbox{\Pifont{psy}r} {\rm X} \right)^2 =
\big< {\rm X}^2 \big> - \big< {\rm X} \big>^2
\end{equation}
and a standard $\Delta{\rm X}$, where
$\mbox{\Pifont{psy}r}$ is the projector or density operator that
describes all the probability distributions of the problem
in question.  Moreover, for two observables represented by
two non-commuting operators A and B, that is,
{\renewcommand{\theequation}{A-9}
\begin{equation}
{\rm AB} - {\rm BA} = i{\rm C}
\end{equation}
it is readily shown that $\Delta{\rm A}$ and
$\Delta{\rm B}$ satisfy
the uncertainty relation [1,9,35]
{\renewcommand{\theequation}{A-10}
\begin{equation}
\Delta {\rm A} \Delta {\rm B} \geq | \big< {\rm C} \big> |/2
\end{equation}

It is evident that each uncertainty relation refers neither to
any errors introduced by the measuring instruments nor to any
particular value of a measurement result.  The reason for
the latter remark is that the value of an observable is determined
by the expectation value of the operator representing the
observable and not by any individual measurement result.

A XVI: {\it Collapse of the wave function postulate}. 
Among the postulates of quantum mechanics, many authoritative
textbooks include von Neumann's projection or collapse of
the wave function postulate [33,36].  As we indicate earlier,
an excellent, rigorous, and complete discussion of the 
falsity of the 
projection
postulate is given by Park [16].

The only thing that we wish to add here is an argument against
the postulate based on a violation of the position-momentum
uncertainty relation.  We consider a structureless particle
confined in a one-dimensional, infinitely deep potential well
of width $L$.  Initially, the particle is in a state
characterized by a projector $|\Psi\big>\big<\Psi|$.
According to the projection postulate, upon a momentum
measurement the particle must collapse into a momentum
eigenstate.  Suppose that the eigenstate just cited is
characterized by the ith momentum projector 
$|{\rm p}_\i\big>\big<{\rm p}_\i|$.  For such a 
projector, the standard deviation of position measurement
results satisfies the relations
$0 < \Delta {\rm x} < L$, and the
standard deviation of momentum measurement results
$\Delta {\rm p} = 0$.  Accordingly, 
$\Delta{\rm x}\Delta{\rm p} = 0 < \hbar/2$ instead of
$\Delta{\rm x}\Delta{\rm p} \geq \hbar/2.$  In view
of the unquestionable validity of uncertainty relations,
we must conclude that the projection postulate cannot be
a valid postulate of quantum theory.

\subsection*{Dynamics: Evolution of the density operator in time}

A XVII: {\it Dynamical postulate}.  Hatsopoulos and
Gyftopoulos [8] postulate that unitary transformations
of $\mbox{\Pifont{psy}r}$ in time obey the relation
{\renewcommand{\theequation}{A-11}
\begin{equation}
{d \mbox{\Pifont{psy}r} \over dt } = - { i \over \hbar}
\left[ {\rm H} \mbox{\Pifont{psy}r} -
\mbox{\Pifont{psy}r} {\rm H} \right]
\end{equation}
where H is the Hamiltonian operator of the system.  
If H is independent of $t$, 
the
unitary transformation of $\mbox{\Pifont{psy}r}$ satisfies
the equation
{\renewcommand{\theequation}{A-12}
\begin{equation}
\mbox{\Pifont{psy}r} (t) = {\rm U} (t, t_0) 
\mbox{\Pifont{psy}r} (t_0) {\rm U}^+ (t,t_0)
\end{equation}
where U$^+$ is the Hermitian conjugate of U
{\renewcommand{\theequation}{A-13}
\begin{equation}
{\rm U} (t,t_0) = \exp[-(i/\hbar)(t-t_0){\rm H}]
\end{equation}
and if H is explicitly dependent on $t$ 
{\renewcommand{\theequation}{A-14}
\begin{equation}
{ d {\rm U} \left( t, t_0 \right) \over dt } = 
- \left( i/\hbar \right) {\rm H} \left( t \right)
{\rm U} \left( t,t_0 \right) .
\end{equation}
}
Though Eq. (A-11) is well known in the literature as the 
von Neumann equation, here it must be postulated for the
following reason. In statistical quantum mechanics [37]
the equation is derived as a statistical average of 
Schr\"odinger equations, each of which describes the evolution
in time of a projector 
$\mbox{\Pifont{psy}r}_\i$ and each of which is multiplied
by a time independent statistical probability $\alpha_\i$.
But here, $\mbox{\Pifont{psy}r}$ is not a mixture of projectors
and therefore cannot be derived as a statistical average of
projectors.  It is noteworthy that the dynamical postulate
(Eq. A-11) is limited or incomplete because all unitary
evolutions of $\mbox{\Pifont{psy}r}$ in time correspond
to reversible adiabatic processes, but not all
reversible adiabatic processes correspond to unitary
evolutions of $\mbox{\Pifont{psy}r}$ in time [8] and
not all processes are reversible.

A nonlinear equation of motion 
that describes both all reversible processes
and all irreversible processes has been conceived by
Beretta et al [24,25].  It is not discussed here for the
sake of brevity.  The only thing we wish to emphasize
is that the Berettta equation is shown to satisfy all
the requirements for it to be a bonafide equation
of motion of a nonstatistical unified theory of quantum
mechanics and thermodynamics [38].

\vfill\eject
\noindent
{\bf References}

\begin{enumerate}
\item M.A. Nielsen and I.L. Chuang, {\it Quantum Computation
and Quantum Information}, Cambridge University Press,
Cambridge, United Kingdom (2000).
\item A. Einstein, B. Podolsky, and N. Rosen, Phys. Rev., 
{\bf 47} 777 (1935).
\item E. Schr\"odinger, Die Naturewissenschaften, {\bf 23}
807; 823; 844 (1935).
\item E. Schr\"odinger, translation in English of Ref. [3],
published in Proc. Amer. Phil. Soc., {\bf 124},
323 (1980).
\item E. Schr\"odinger, Proc. Cambridge Phil. Soc., {\bf 31},
555 (1935). Presented by Born.
\item E. Schr\"odinger, Proc. Cambridge, Phil. Soc.,
446 (1936).  Presented by Dirac.
\item J. von~Neumann, {\it Mathematical Foundations of
Quantum Mechanics}, Princeton University Press, NJ (1955).
\item G.N. Hatsopoulos, and E.P. Gyftopoulos, Found. of
Physics, 6, 15; 6, 127; 6, 439; 6, 561 (1976).
\item J.M. Jauch, {\it Foundations of Quantum Mechanics},
Addison-Wesley, Reading, MA (1968).
\item T.S. Kuhn, {\it The Structure of Scientific Revolutions},
2nd Edition, Chicago University Press, Chicago (1970).
\item E.P. Gyftopoulos, Int. J. Therm. Sci. {\bf 38} 741 (1999).
\item E.P. Gyftopoulos, M.R. von Spakovsky, J. Energy Resources
Tech. {\bf 125}, 1 (2003).
\item H. Margenau, {\it The Nature of Physical Reality}, 
Oxbow Press, Woodbridge, Conn. (1977).
\item H. Margenau, Phys. Rev. {\bf 49} 240 (1936).
\item H. Margenau, Phil. of Sci. {\bf 30}, 6 (1963).
\item J.L. Park, Phil. of Sci. {\bf 35} 3 (1968).
\item J.L. Park, AJP, {\bf 211} (1968).
\item R.C. Tolman, {\it The Principles of Statistical Mechanics},
Oxford University Press, Amen House, London (1962).
\item B.M. Terhal, M.M. Wolf, A.C. Doherty, Physics Today,
April 2003, p. 46.
\item J.S. Bell, Physics {\bf 1}, 195 (1964).
\item R. Josza, N. Linden, arXiv:quant-ph/0201143, v2,
8 Mar 2002.
\item A. W\'ojcik, Sci. 301, 5637, 1183 (2003).
\item J.L. Park, N. Margenau, Int. J. Theor. Phys.,
{\bf 1} 211 (1968).
\item G.P. Beretta, E.P. Gyftopoulos, J.L. Park, and
G.N. Hatsopoulos, 
{\it Nuovo Cimento}, 82B, 2, 169-191 (1984).
\item G.P. Beretta, E.P. Gyftopoulos, and J.L. Park,
{\it Nuovo Cimento}, 
87B, 1, 77-97 (1985).
\item E.T. Jaynes, Phys. Rev. 108(2) (1957).
\item A. Katz, {\it Principles of Statistical Mechanics,
The Information Theory Approach}, W.H. Freeman and
Company, San Francisco, CA (1967).
\item E.P. Gyftopoulos and E. Cubukcu, Phys. Rev. E,
55(4) (1997).
\item E.P. Gyftopoulos, J. Energy Res. and Tech.,
Trans ASME, 123 (2001).
\item G.C. Wick, A.S., Wightman, and E.P. Wigner, Phys. 
Rev. 88(1) (1952).
\item W. Heisenberg, {\it The Principles of the Quantum
Mechanics}, translated into English by C. Eckart and
F.C. Hoyt, Dover Publications (1930).
\item W.H. Louisell, {\it Quantum Statistical Properties
of Radiation}, John Wiley \& Sons, New York, NY (1973).
\item R. Shankar, {\it Principles of Quantum Mechanics},
Plenum Press, New York, NY (1994).
\item W.E. Lamb and R.C. Rutherford, Phys. Rev. 72(3)
(1947).
\item D.J. Griffiths, {\it Introduction to Quantum
Mechanics}, Prentice Hall, Englewood Cliffs, NJ (1994).
\item R.L. Liboff, {\it Introductory Quantum Mechanics},
Addison-Wesley, Reading, MA (1980).
\item R.C. Tolman, {\it The Principles of Statistical
Mechanics}, Oxford University Press, Amen House, London
(1962).
\item H.J. Korsh and H. Steffen, J. of Phys. 
A20, 3787 (1987).
\end{enumerate}







\end{document}